\documentclass[conference]{IEEEtran}
\IEEEoverridecommandlockouts
\usepackage{cite}
\usepackage{amsmath,amssymb,amsfonts}
\usepackage{algorithmic}
\usepackage{graphicx}
\usepackage{textcomp}
\usepackage{xcolor}
\usepackage{subfig}
\usepackage{setspace}
\usepackage{algorithm}
\usepackage{setspace}
\usepackage{float}
\usepackage{epstopdf}
\usepackage{makecell}
\usepackage{color}
\usepackage{bbm}
\newtheorem{definition}{Definition}
\newtheorem{challenge}{Challenge}

\usepackage{geometry}
\usepackage[colorlinks,linkcolor=blue]{hyperref}

\def\BibTeX{{\rm B\kern-.05em{\sc i\kern-.025em b}\kern-.08em
    T\kern-.1667em\lower.7ex\hbox{E}\kern-.125emX}}
\begin{document}

\title{Multi-objective Deep Reinforcement Learning for Mobile Edge Computing}

\author{Ning Yang, 
        Junrui Wen, 
        Meng Zhang{*}, 
        Ming Tang 

\thanks{Ning Yang and Junrui Wen are with Institute of Automation, Chinese Academy of Sciences, Beijing, 100190, China. (e-mail: ning.yang@ia.ac.cn, yvwtogo@gmail.com).

Meng Zhang is with the ZJU-UIUC Institute, Zhejiang University, Zhejiang, 314499, China. (e-mail: mengzhang@intl.zju.edu.cn).

Ming Tang is with the Department of Computer Science and Engineering, Southern University of Science and Technology, Shenzhen, 518055, China. (e-mail: tangm3@sustech.edu.cn).

({*}Corresponding author: Meng Zhang)
}}

\maketitle

\begin{abstract}

Mobile edge computing (MEC) is essential for next-generation mobile network applications that prioritize various performance metrics, including delays and energy consumption. However, conventional single-objective scheduling solutions cannot be directly applied to practical systems in which the preferences of these applications (i.e., the weights of different objectives) are often unknown or challenging to specify in advance. In this study, we address this issue by formulating a multi-objective offloading problem for MEC with multiple edges to minimize expected long-term energy consumption and transmission delay while considering unknown preferences as parameters.
To address the challenge of unknown preferences, we design a multi-objective (deep) reinforcement learning (MORL)-based resource scheduling scheme with \textit{proximal policy optimization} (PPO). In addition, we introduce a well-designed state encoding method for constructing features for multiple edges in MEC systems, a sophisticated reward function for accurately computing the utilities of delay and energy consumption. Simulation results demonstrate that our proposed MORL scheme enhances the hypervolume of the Pareto front by up to $233.1\%$ compared to benchmarks. Our full framework is available at \url{https://github.com/gracefulning/mec_morl_multipolicy}.
\end{abstract}


\section{Introduction}
The rise of next-generation networks and the increasing use of mobile devices have resulted in an exponential growth of data transmission and diverse computing needs. 
With the emergence of new computing-intensive applications, there is a possibility that device computing capacity may not suffice. Cloud computing is one solution that can provide the necessary resources, but it may also result in latency issues.

To address this challenge, mobile edge computing (MEC) has emerged as a promising computing paradigm that offloads computing workload to edge or cloud networks and can achieve low latency and high efficiency \cite{mach2017mobile, mao2016dynamic, you2016energy}.

In MEC systems, task offloading is crucial in achieving low latency and energy consumption \cite{li2018deep}. By selectively offloading computing tasks to edge or cloud users based on their requirements, MEC systems can optimize resource utilization and improve performance. For example, edge servers may be effective for low-latency tasks that require real-time processing, while cloud users may be more suitable for computationally intensive tasks. Additionally, other factors, such as edge load and transmission rate, need to be considered when designing offloading schemes. Task offloading schemes in MEC systems present two key challenges.

\begin{challenge}{\textit{The natural MEC network environments are full of dynamics and uncertainty.} }
\end{challenge}

The scheduling of offloading in MEC systems is challenging due to the dynamic and unpredictable nature of users' workloads and computing requirements. 
The presence of stochastic parameters in the problem poses challenges to the application of traditional optimization methods. 
Myopically optimizing the offloading decision of the current step is ineffective since it cannot account for long-term utilities.

The application of deep reinforcement learning (DRL) has shown substantial potential in addressing sequential decision-making problems and is an attractive technique for dynamic MEC environments \cite{mnih2013playing,li2018deep}. 
The existing works have demonstrated the effectiveness of applying DRL in MEC systems to address unknown dynamics.
For instance, Cui et al. \cite{cui2020latency} employed DRL to solve the user association and offloading sub-problem in MEC networks. Lei et al. \cite{lei2019multiuser} investigated computation offloading and multi-user scheduling algorithms in edge IoT networks and proposed a DRL algorithm to solve the continuous-time problem, supporting implementation based on semi-distributed auctions. Jiang et al. \cite{jiang2020stacked} proposed an online DRL-based resource scheduling framework to minimize the delay in large-scale MEC systems. However, there is another challenge that requires consideration.

\begin{challenge}{\textit{Users who initiate tasks may have diverse preferences regarding delay and energy consumption.} }
\end{challenge}

In various mobile applications such as health care, transportation, and virtual reality, among others, delay in processing data can have serious consequences, particularly in emergency situations. However, in industrial and unmanned aerial networks, energy consumption is subject to strict limits, and thus, computing applications in these areas may prioritize energy over delay. Therefore, offloading scheduling in MEC systems requires a well-designed balance between delay and energy consumption. Moreover, one of the most critical considerations in designing an offloading scheme for MEC systems is that target applications may not be known in advance. 

Regretfully, existing studies on MEC (e.g., \cite{cui2020latency, lei2019multiuser, jiang2020stacked, li2018deep, yang2020partially
, zhang2020power
}), most of them have focused exclusively on single-objective methods. In practice, many scheduling problems in MEC systems are in nature \textit{multi-objective}. Since these studies have not taken into account multi-objective methods, they cannot address the second challenge of MEC systems, which is dealing with diverse and unknown preferences. The dynamic and uncertain nature of the environments, the diversity of preferences, and the computational infeasibility of classical methods motivate us to seek out new methodologies to address 
these issues.



Note that although some may argue that we can still directly apply single-objective DRL by simply taking a weighted sum (known as scalarization), this is, in fact, not true due to the following issues \cite{roijers2013survey}: 
\begin{enumerate}
    \item \textit{Impossibility}: Weights may be unknown when designing or learning an offloading scheme.
    \item \textit{Infeasibility}: Weights may be diverse, which is true when MEC systems have different restrictive constraints on latency or energy.
    \item \textit{Undesirability}: Even if weights are known, nonlinear objective functions may lead to non-stationary optimal policies.
\end{enumerate}

To effectively address these challenges, we propose employing multi-objective reinforcement learning (MORL) to design a task offloading method. We summarize our main contributions as follows:
\begin{itemize}
\item\textit{Multi-objective MEC Framework}: We formulate the multi-objective MDP (Markov decision process) problem framework. Compared with previous works, our framework focuses on the Pareto optimal solutions, which characterize the performance of the offloading scheduling policy with multiple objectives under different preferences.

\item \textit{Multi-objective Decision Model}: We propose a novel MORL method based on proximal policy optimization (PPO) to solve the multi-objective problem. Our proposed method aims to achieve the Pareto near-optimal solution for diverse preferences. Moreover, we introduce a well-designed encoding method to construct features for multi-edge systems and a sophisticated reward function to compute delay and energy consumption.

\item \textit{Numerical Results:} Compared to benchmarks, our MORL scheme increases the hypervolume of the Pareto front up to $233.1\%$.

\end{itemize}

\section{System Model}

\begin{figure}[t]
\small
\centering
\includegraphics*[width=70mm]{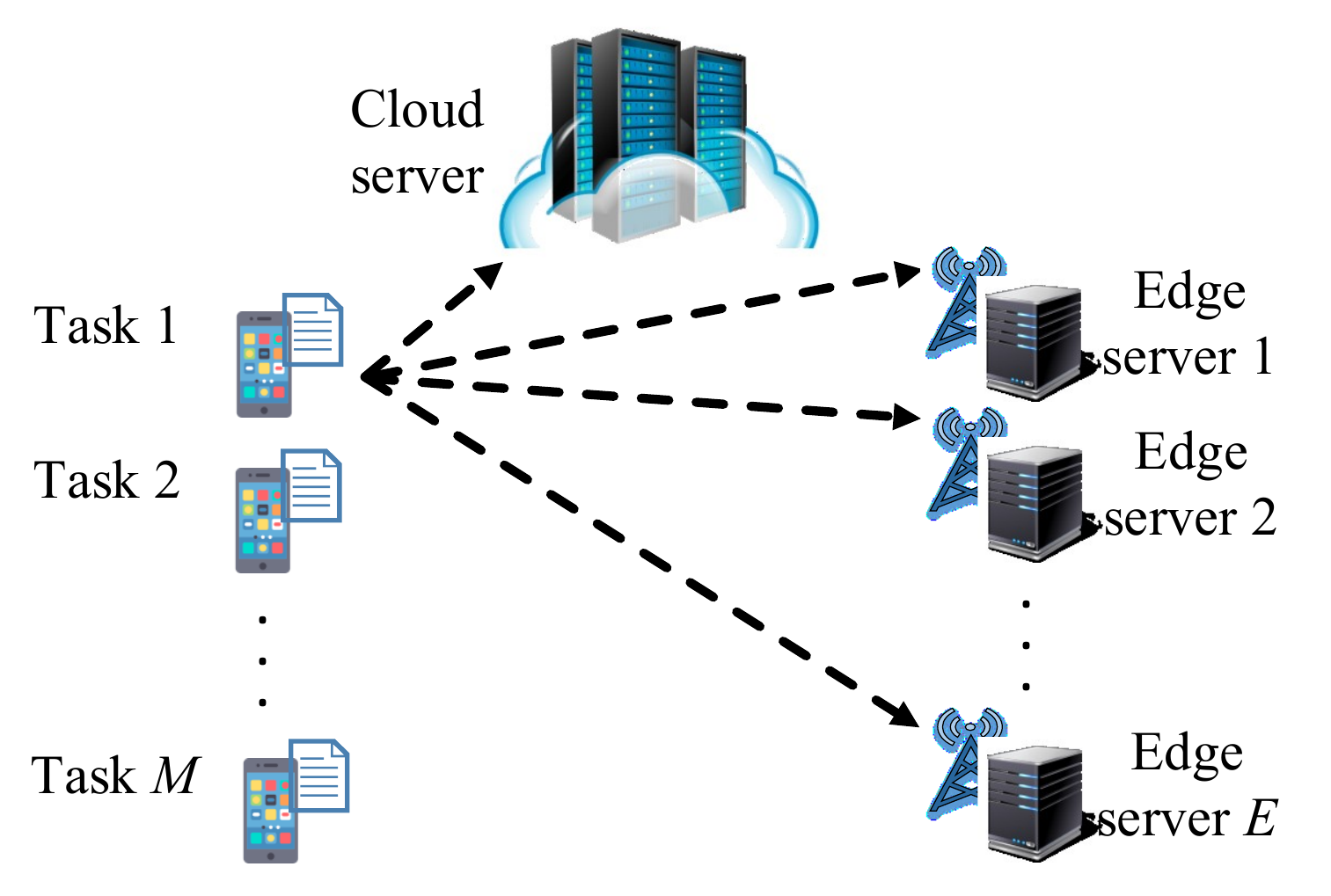}
\caption{An illustrative example system model of MEC.}
\label{fig:System}
\end{figure}

We consider a set of servers  $\mathcal{E}=\{0,1,2,...,E\}$ with one remote cloud server (denoted by index $0$) and $E$ edge servers, and consider a set of users $\mathcal{U}=\{1,2,...,U\}$ in a MEC system, as shown in Fig. \ref{fig:System}.
We use index $e \in \mathcal{E}$ to denote a server. Index $u \in \mathcal{U}$ denotes a user.
Our model is a continuous-time system and has discrete decision steps. Consider one episode consisting of $T$ steps, and each step is denoted by $t \in \{ 1,2,...,{T} \}$, each with a duration of $\Delta t$ seconds.

\enlargethispage{-1pt}
Multiple users request MEC services from servers.
At the beginning of each step, the arrival time of a series of tasks follows a Poisson distribution for each user, and the Poisson arrival rate for each user is $\lambda_p$. The tasks are placed in a queue with a first in, first out (FIFO) queue strategy.
In each step, the system will offload the first task in the queue to one of the servers. Then the task is removed from the queue. Let $\mathcal{M} = \{1,2,..., M\}$ denote the set of tasks in an episode. We use $m \in \mathcal{M}$ to denote a task and use ${L_m}$ to denote the size of task $m$, which follows an exponential distribution \cite{lei2019joint} with mean $\bar L$.
\vspace{0pt}

We consider a Rayleigh fading channel model in the MEC network. We denote $\boldsymbol{h} \in \mathbb{R}^{U \times (E+1)}$ as the $U\times (E+1)$ channel matrix. Thus, the achievable data rate from user $u$ to server $e$ is
\begin{equation}
{{C}_{u,e} = {W{{\log }_2}\left(1 + \frac{p^{\rm off}{|h_{u,e}|}^2}{\sigma^2}\right)}}, \forall u\in\mathcal{U}, e\in\mathcal{E},
\end{equation}
where ${\sigma ^2}$ is additive white Gaussian noise (AWGN) power, and $W$ is the bandwidth. The offloading power is $p^{\rm off}$, and the channel coefficient from user $u$ to server $e$ is $h_{u,e}$.

\textbf{Offloading:}
We denote the offloading decision (matrix) as $\boldsymbol{x}=\{x_{m,e}\}_{m\in\mathcal{M},e\in\mathcal{E}} $, where $x_{m,e} \in \{0, 1\}$ is an offloading indicator variable; $x_{m,e}=1$ indicates that task $m$ is offloaded to server $e$. If task $m$ comes from user $u$. The offloading delay for task $m$ is given by \cite{2020EnergyConsumption}
\begin{equation}
T_m^{{\rm{off}}} = \sum\limits_{e \in \mathcal{E}}{x_{m,e}}\frac{L_m}{C_{u,e}},  
~~\forall m\in\mathcal{M}.
\label{eq:offloading delay}
\end{equation}
The offloading energy consumption for task $m$ with offloading power  $p^{\rm off}$ is
\begin{equation}
E_m^{{\rm{off}}} = p^{{\rm{off}}}T_m^{{\rm{off}}},~~\forall m\in\mathcal{M}.
\end{equation}

\textbf{Execution:}
Each server executes tasks in parallel. We denote the beginning of step $t$ as time instant $\tau_t$, given by $\tau_t=t\Delta t$. The computing speed for each task in server $e$ at time instant $\tau_t$ is
\begin{equation}
{q_{e}(\tau_t) = {\frac{f_{e}}{n_{e}^{\rm exe}(\tau_t)\eta}}},~~\forall e\in\mathcal{E},
\end{equation}
\enlargethispage{-5pt}
where $f_e$ is the CPU frequency (in cycles per second) of server $e$, and $\eta$ is the number of CPU cycles required for computing a one-bit task. We define $n_e^{\rm exe}(\tau_t)$ as the number of tasks that are being executed in server $e$ at time $\tau_t$. The $n_e^{\rm exe}(\tau_t)$ tasks share equally the computing resources of server $e$. Thus, we give the relation between task size $L_m$ and execution delay $T_m^{\rm exe}$ for task $m$ as 
\begin{equation}
\begin{array}{l}
\begin{aligned}
{L_m} & = g_{m}(T_m^{\rm exe})\\ 
&~{= \sum\limits_{e \in \mathcal{E}}x_{m,e} \int_{{m\Delta t}+T_m^{\rm off}}^{m{\Delta t}+T_m^{\rm off}+T_m^{\rm exe}} {q_{e}(\tau)}\, d\tau},\forall m\in\mathcal{M},
\end{aligned}
\end{array}
\end{equation}
where $\tau$ is a time instant. The integral function $g_{m}(T_m^{\rm exe})$ denotes the aggregate executed size for task $m$ from $m{\Delta t}+T_m^{\rm off}$ to $m{\Delta t}+T_m^{\rm off}+T_m^{\rm exe}$. Therefore,  execution time delay $T_m^{\rm exe}$ of task $m$ is
\begin{equation}
{T_m^{\rm exe}={g_m}^{-1}(L_m)},
\forall m\in\mathcal{M}.
\end{equation}
The total energy consumption of execution for task $m$ is given by \cite{2020EnergyConsumption}
\begin{equation}
\begin{array}{l}
\begin{aligned}
{E_m^{\rm exe} =\sum\limits_{e \in \mathcal{E}} {x_{m,e}} {\kappa{\eta}f_e^2 {L_m}}},\forall m\in\mathcal{M}, 
\end{aligned}
\end{array}
\end{equation}
where $\kappa$ denotes an effective capacitance coefficient for each CPU cycle. 

To summarize, the overall delay and the overall energy consumption for task $m\in\mathcal{M}$ are
\begin{align}
{T_m}= T_m^{{\rm off}} + T_m^{\rm exe},{E_m}= E_m^{\rm off} + E_m^{\rm exe},
\label{eq:execution energy consumption for user m}
\end{align}
respectively. 

The mean of task size $\bar{L}$ represents the demand for tasks. If the computational capability of the system exceeds the demand, the scheduling pressure decreases. Conversely, if the demand surpasses the capability, the system will continuously accumulate tasks over time. Therefore, we consider a system that balances computational capability and task demand.
The mean of task size $\bar L$ satisfies

\begin{equation}
{\Delta t\left(\sum \limits_{e \in \mathcal{E}} \frac{f_e}{\eta}\right)=\lambda_p \bar {L} U},
\label{eq:Balance}
\end{equation}
.
\subsection{Problem Formulation}\label{subsec:Problem Formulation}
Based on different application scenarios, MEC networks have diverse preferences over energy consumption and delay. Therefore, we aim to design a scheduling policy to achieve the Pareto optimal solution between energy consumption and delay. We cannot directly apply single-objective DRL by simply taking a weighted sum due to impossibility (i.e., weights may be unknown), infeasibility (i.e., MEC systems have different restrictive constraints on latency or energy), and undesirability (i.e., non-stationary optimal policies). This motivates us to use MORL to achieve Pareto optimal solution for any potential preference. We introduce 
the preference vector $\boldsymbol{\omega}=(\omega_{\rm T},\omega_{\rm E})$ to weight delay and energy consumption, which  satisfies $\omega_{\rm T} + \omega_{\rm E}=1$. The subscript $\rm T$ denotes delay about, while the subscript $\rm E$ denotes energy consumption about in our study.

A (stochastic) policy is a mapping $\pi: \mathcal{S} \times \mathcal{A} \to [0,1]$, where $\mathcal{S}$ is the state space of the system and $\mathcal{A}$ is the offloading action space, we will formally define them in next section. For any given task $m$ and system state, policy $\pi$ selects an offloading decision $x_{m,e}$ according to a certain probability distribution. 
Given any one possible $\boldsymbol{\omega}$, the multi-objective resource scheduling problem under the policy $\pi$ is given by
\begin{subequations}\label{eq:Optimization problem}
\begin{align}
\min_{\pi} &\quad \mathbb{E}_{\boldsymbol{x} \sim \pi} \left[ \sum_{m\in\mathcal{M}} \gamma^m \left(\omega_{\rm T} T_m+\omega_{\rm E} E_m \right)\right]
\label{eq:Optimization function}\\
{\rm s.t.}& \quad x_{m,e} \in \{ 0,1\},~~\forall m\in\mathcal{M},\forall e\in\mathcal{E}, \label{eq:Model constraints1}\\
&{\quad \sum_{e\in\mathcal{E}} {x}_{m,e} \leq 1,~~\forall m\in\mathcal{M},}
\label{eq:Model constraints2}
\end{align}
\end{subequations}
where constraint \eqref{eq:Model constraints1} restricts task offloading variables to be binary, and constraint \eqref{eq:Model constraints2} guarantees that each task can be only offloaded to one server. A discount factor $\gamma$ characterizes the discounted objective in the future. The expectation $\mathbb{E}$ accounts for the distribution of the task size $L_m$, the arrival of users , and stochastic policy $\pi$.

\subsection{Multi-objective Metrics}\label{subsec:Multi-objective metrics}
To facilitate multi-objective analysis, we further introduce the following notions. Consider a preference set $\Omega=\{\boldsymbol\omega_1,\boldsymbol \omega_2,...,\boldsymbol \omega_n\}$ with $n$ preferences.
A scheduling policy set $\Pi=[\pi_1,\pi_2,...,\pi_n]$ with $n$ policies solving problem \eqref{eq:Optimization function} given corresponding preferences in $\Omega$.
Let $\boldsymbol{y}$ denote the performance, given by
\begin{equation}
\boldsymbol{y}=\{y_{\rm T},y_{\rm E}\}=\left\{\sum_{m\in\mathcal{M}}T_m,\sum_{m\in\mathcal{M}}E_m\right\}.
\end{equation}
A performance of $\Pi$ is denoted as $\boldsymbol{Y}=\{\boldsymbol{y}^{\pi_1},\boldsymbol{y}^{\pi_2},...,\boldsymbol{y}^{\pi_n}\}$. We consider the following definition to characterize the optimal trade-offs between two performance metrics:
\begin{definition}[Pareto front \cite{roijers2013survey}]{\emph{For a policy set $\Pi$, Pareto front $PF(\Pi)$ is the undominated set :
\begin{equation}
{PF(\Pi)=\{\pi \in \Pi~|~\nexists \pi^{\prime}\in\Pi:\boldsymbol{y}^{\pi^{\prime}} \succ_P \boldsymbol{y}^{\pi}} \},
\label{def:Pareto front}
\end{equation}}}
\end{definition}

\emph{where $\succ_P$ is the Pareto dominance relation, satisfying
\begin{equation}
\begin{array}{l}
\boldsymbol{y}^{\pi} \succ_P \boldsymbol{y}^{\pi^{\prime}} \iff \\ (\forall i : y_{i}^{\pi} \ge y_{i}^{\pi^{\prime}}) \land (\exists i : y_{i}^{\pi} > y_{i}^{\pi^{\prime}}), i \in \{{\rm T},{\rm E}\}.
\end{array}
\end{equation}}
We aim to approximate the exact Pareto front\cite{roijers2013survey} by searching for policies set $\Pi$.
The following hypervolume metric can measure the quality of an approximation:
\begin{definition}[Hypervolume metric \cite{zitzler1999multiobjective}]{\emph{In the multi-objective MEC scheduling problem, as a Pareto front approximation $PF(\Pi)$, the hypervolume metric is}}
\begin{equation}
\begin{array}{l}
\mathcal{V}(PF(\Pi))=\int_{\mathbb{R}^2} {\mathbb{I}_{V_h(PF(\Pi))}(z)dz},
\end{array}
\label{def:Hypervolume metric}
\end{equation}
\end{definition}
where $V_h(PF(\Pi))=\{z \in Z| \exists \pi \in PF(\Pi) : \boldsymbol{y}^{\pi} \succ_P z \succ_P \boldsymbol{y}^{\rm ref}\}$, and $\boldsymbol{y}^{\rm ref} \in \mathbb{R}^2$ is a reference performance point. Function $\mathbb{I}_{V_h(PF(\Pi))}$ is an indicator function that returns $1$ if $z \in V_h(PF(\Pi)^{\prime})$ and $0$ otherwise.

The multi-objective resource scheduling problem is still a challenge for MEC networks for the following reasons:

\begin{itemize}
\item The natural MEC network environments are full of dynamics and uncertainty, leading to unknown preferences of MEC systems.

\item The computation complexity of the conventional optimization method is demanding since the goal is to get a vector reward instead of a reward value. The objective function \eqref{eq:Optimization function} and the feasible set of constraints \eqref{eq:Model constraints1} and \eqref{eq:Model constraints2} are non-convex due to binary variables $\boldsymbol {x}$.

\end{itemize}

The aforementioned problems motivate us to design a MORL scheme to solve \eqref{eq:Optimization problem}.

\section{MORL Scheduling Solution}
This section considers the situation of multiple preferences. We consider that a (central) agent makes all offloading decisions in a fully-observable setting.
We model the MEC environment as a MOMDP framework. In the subsection, we first introduce the MOMDP framework, which includes a well-designed state encoding method and a sophisticated reward function. Then, we present our algorithm by introducing aspects including the neural network architecture and policy update method.

\vspace{-0pt}
\subsection{The MOMDP Framework}\label{subsec:The MOMDP Framework}
\begin{definition}[MOMDP \cite{roijers2013survey}]{\emph{
A MOMDP  is a tuple $\langle \mathcal S, \mathcal A, \mathcal T, \gamma, \mu, \mathcal R \rangle$ that contains state space $\mathcal  S$, action space $\mathcal  A$, probabilistic transition process $ \mathcal {T:  S \times  A \to  S} $,  discount factor $\gamma \in [0, 1)$, a probability distribution over initial states $\mu :\mathcal  S \to [0, 1]$, and a vector-valued reward function $\mathcal {R: S \times A} \to\mathbb{R}^2$ that specifies the immediate reward for the delay objective and the energy consumption objective.}}
\end{definition}


For a decision step $t$, an agent offloads task $m$ from user $u$. It has $m=t$ for task index $m$ and step-index $t$. We specify the MOMDP framework in the following:

{\bf{State $\mathcal S$}}: We consider $E+1$ servers ($E$ edge servers and a cloud server). Hence, the state $\boldsymbol{s}_t \in {\mathcal S}$ at step $t$ is a fixed length set and contains $E+1$ server information vectors. We formulate state $\boldsymbol s_{t}$ as $\boldsymbol{s}_t=\{ \boldsymbol s_{t,e} | e \in \mathcal{E} \}$. 
The information vector of server $e$ at step $t$ is
\begin{equation}
\boldsymbol s_{t,e} =(L_m,{C}_{u,e},f_e,n_e^{\rm exe}(\tau_t),E,\boldsymbol {\mathcal B}_{e}),~~~ \forall e \in \mathcal{E}.
\end{equation}
State $\boldsymbol s_{t,e}$ contains task size ${L_m}$, data rate ${{C}_{u,e}}$, CPU frequency $f_e$, the number of execution task ${n_e^{\rm exe}}(\tau_t)$, the number of edge server $E$, and task histogram vector $\boldsymbol {\mathcal B}_{e}$, which is the residual size distribution for tasks executed in server $e$ at time instant $\tau_t$. That is,

\vspace{-10pt}
\begin{equation}
\boldsymbol {\mathcal B}_e(\tau_t)=(b_{1,e}^{\rm exe}(\tau_t),b_{2,e}^{\rm exe}(\tau_t),...,b_{N,e}^{\rm exe}(\tau_t)).
\label{eq:histogram vector}
\end{equation}
Histogram vector $\boldsymbol {\mathcal B}_{e}$ has $N$ bins. We denote one of previous tasks as $m^{\prime}$ and denote the execution residual size of task $m^{\prime}$ at time instant $\tau_t$ as $L_{m^{\prime}}^{\rm res}(\tau_t)$.
In Eq. \eqref{eq:histogram vector}, the $i$-th value $b_{i,e}^{\rm exe}(\tau_t)$ in $\boldsymbol {\mathcal B}_{e}$ denotes the number of tasks with execution residual size $L_{m^{\prime}}^{\rm res}(\tau_t)$ within the range of $[i-1,i)$ Mbits. 
Specifically, the last element $b_{N,e}^{\rm exe}(\tau_t)$ denotes the number of tasks with execution residual size $L_{m^{\prime}}^{\rm res}(\tau_t)$ within the range of $[N-1,+\infty)$ Mbits. 
The execution residual size $L_{m^{\prime}}^{\rm res}(\tau_t)$ 
is given by
\begin{equation}
\begin{array}{l}
L_{m^{\prime}}^{\rm res}(\tau_t) = L_{m^{\prime}}- {\rm min}\left( g_{{m^{\prime}}}\left((\tau_t-{m^{\prime}} \Delta t \right),L_{m^{\prime}} \right),
\\
~~~~~~~~~~
\forall \tau_t \in [t\Delta t,T\Delta t],m^{\prime} \in \{1,2,\dots,m-1\}.
\end{array}
\end{equation} 

\enlargethispage{-2pt}
{\bf{Action $\mathcal A$}}: The action $a_t \in \mathcal A$ denotes that offloading task $m$ to which server. The action space is $\mathcal{A}=\{0,1,2,\dots,E\}$. Hence, the action at step $t$ is represented by the following

\begin{equation}
{a_t = \sum \limits_{e \in \mathcal{E}} e{x}_{m,e}(t)}.
\end{equation}


{\bf{Transition $\mathcal T$}}:
It describes the 
transition from $\boldsymbol s_t$ to $\boldsymbol {s}_{t+1}$ with action $a_t$, which is denoted by $P({\boldsymbol{s}_{t + 1}}|{\boldsymbol{s}_t},{a_t})$.

{\bf{Reward $\mathcal {R}$}}: Unlike a classical MDP setting in which each reward is a scalar, a multi-objective setting requires a vector. Therefore, our reward (profile) function is given by $\mathcal {R: S \times A} \to\mathbb{R}^2$. 
We denote the reward of energy consumption and delay as ${r}_{\rm E}$ and ${r}_{\rm T}$. If the agent offloads task $m$ to server $e$ at step $t$, the reward of energy consumption for state $\boldsymbol{s}_t$ and action $a_t$ is 
\begin{equation}
{{{r}_{\rm E}}(\boldsymbol{s}_t,a_t) = -{\hat E}_{m} },
\label{eq:Reward of energy consumption}
\end{equation}
where ${\hat E}_{m}$ is the estimated energy consumption of task $m$. Through \eqref{eq:execution energy consumption for user m}, we can compute the energy consumption of task $m$. The MORL algorithm maximizes the reward, which is thus the negative of energy consumption. For one episode, the total reward for energy consumption is given by
\begin{equation}
R_{\rm E}=\sum \limits_{t=1}^{T} r_{\rm E}(\boldsymbol{s}_t,a_t) = -\sum \limits_{m \in \mathcal{M}} \hat{E}_{m}.
\label{eq:Total reward of energy consumption}
\end{equation}
The reward for the delay is
\begin{equation}
{r}_{\rm T}(\boldsymbol{s}_t,a_t) = -\left({\hat T}_{m} + \sum \limits_{m^{\prime} \in \mathcal{M}_{e}(\tau_t)} \Delta {\hat T}_{m^{\prime}}^{a_t}\right),
\label{eq:Reward of delay}
\end{equation}
where ${\hat T}_{m}$ is the estimated delay for task $m$, and $\mathcal{M}_{e}(\tau_t)$ is a set of tasks, which are executed in server $e$ at time instant $\tau_t$. The estimated correction of delay $\Delta {\hat T}_{m^{\prime}}^{a_t}$ describes how much delay will increase to task $m^{\prime}$ with action $a_t$. For one episode, the total reward of delay has
\begin{equation}
R_{\rm T}=\sum \limits_{t=1}^{T} r_{\rm T}(\boldsymbol{s}_t,a_t) = -\sum \limits_{m \in \mathcal{M}} T_m.
\label{eq:Total reward of delay}
\end{equation}
To compute reward $r_T$, we rewrite Eq.\eqref{eq:Reward of delay} as
\begin{equation}
\begin{split}
{{r}_{\rm T}}(\boldsymbol{s}_t,a_t) =- {\hat T}_{m} - \sum \limits_{m^{\prime} \in \mathcal{M}_e(\tau_t)} ({\hat T}_{m^{\prime}}^{a_t} - {\hat T}_{m^{\prime}}^{a^*(t)}),
\label{eq:Reward of delay1}
\end{split}
\end{equation}
where ${\hat T}_{m^{\prime}}^{a_t}$ denotes the estimated residual delay of task $m^{\prime}$ with taking action $a_t$ at step $t$. The residual delay of task $m^{\prime}$ without taking action $a_t$ is ${\hat T}_{m^{\prime}}^{a^*(t)}$, which is the estimated residual delay at the end of step $t-1$. Next, we introduce the computation of the two cases.

\enlargethispage{-4pt}
(1) \textit{The case without taking action $a_t$}:
For task set ${\mathcal{M}_{e}(\tau_t)}$ with $n_{e}^{\rm exe}(\tau_t)$ tasks, the execution residual size is a set $\mathcal{L}_{\mathcal{M}_{e}(\tau_t)}^{\rm res} = \{ L_{m^{\prime}}^{\rm res}(\tau_t) | {m^{\prime}} \in  \mathcal{M}_{e}(\tau_t)\}$. We sort residual task size set $\mathcal{L}_{\mathcal{M}_{e}(\tau_t)}^{\rm res}$ in the ascending order and get a vector $\boldsymbol{L}_{\mathcal{M}_{e}(\tau_t)}^{\rm sort}=(L_{1,e}^{\rm sort}(\tau_t),L_{2,e}^{\rm sort}(\tau_t),...,L_{n_{e}^{\rm exe}(\tau_t),e}^{\rm sort}(\tau_t))$, where $L_{i,e}^{\rm sort}(\tau_t)$ is the $i$-th least residual task size in $\mathcal{L}_{\mathcal{M}_{e}(\tau_t)}^{\rm res}$. Specifically, we define 
$L_{0,e}^{\rm sort}(\tau_t)=0$. Then, we have 
\begin{equation}
\begin{split}
~~~~~\sum \limits_{{m}^{\prime} \in \mathcal{M}_e(\tau_t)} {\hat T}_{{m}^{\prime}}^{a^*(t)}
=\!\!\! \sum \limits_{i=1}^{n_{e}^{\rm exe}(\tau_t)} (n_{e}^{\rm exe}(\tau_t)\!\!-i\!\!+1) {\hat T}_{i,e}^{\rm dur}
~~~~~~~~~~~~~~~~~\\
=\!\!\!\sum \limits_{i=1}^{n_{e}^{\rm exe}(\tau_t)}\!\!\! \frac{\eta}{f_e}(n_{e}^{\rm exe}(\tau_t) -i\!+1)^2 (L_{i,e}^{\rm sort}(\tau_t) - L_{i-1,e}^{\rm sort}(t)),~~~~~~~~~~~
\label{eq:Reward of delay A}
\end{split}
\end{equation}
where ${\hat T}_{i,e}^{\rm dur}$ denotes the estimated during of time from the completing instant of residual task $L_{i-1,e}^{\rm sort}(\tau_t)$ to the completing instant of residual task $L_{i,e}^{\rm sort}(\tau_t)$.

(2) \textit{The case with action $a_t$}:
The MEC system completes offloading task $m$ at time instant
$\tau_t^{\prime} = \tau_t + T_m^{\rm off}$. We consider a high-speed communication system that offloading delay $T_m^{\rm off}$ is short than the duration of one step $\Delta t$ and satisfies $T_m^{\rm off} < \Delta t $.
For task set 
${\mathcal{M}_{e}(\tau_t^{\prime})}$ 
with $n_{e}^{\rm exe}(\tau_t^{\prime})$
tasks, the execution residual size is a set $\mathcal{L}_{\mathcal{M}_{e}(\tau_t^{\prime})}^{\rm res} = \{ L_m^{\rm res}(\tau_t^{\prime}) | m \in  \mathcal{M}_{e}(\tau_t^{\prime})\}$. 
We sort 
set $\mathcal{L}_{\mathcal{M}_{e}(\tau_t^{\prime})}^{\rm res}$ 
in the ascending order and get a vector $\boldsymbol{L}_{\mathcal{M}_{e}(\tau_t^{\prime})}^{\rm sort}=(L_{1,e}^{\rm sort}(\tau_t^{\prime}),L_{2,e}^{\rm sort}(\tau_t^{\prime}),...,L_{n_{e}^{\rm exe}(\tau_t^{\prime}),e}^{\rm sort}(\tau_t^{\prime}))$, where $L_{i,e}^{\rm sort}(\tau_t^{\prime})$ 
is the $i$-th least residual task size in $\mathcal{L}_{\mathcal{M}_{e}(\tau_t^{\prime})}^{\rm res}$. 
Then, it satisfies 
\begin{equation}
\begin{split}
{\hat{T}_m + \sum \limits_{m^{\prime} \in \mathcal{M}_e(\tau_t^{\prime})} {\hat T}_{m^{\prime}}^{a_t}}
~~~~~~~~~~~~~~~~~~~~~~~~~~~~~~~~~~~~~~~~~~~~~~~
\\ 
= \!\!\sum \limits_{i=1}^{n_{e}^{\rm exe}(\tau_t)} \!\!(n_{e}^{\rm exe}\!\!-\!i\!+\!1) {\rm min}\!\!
\left( 
{\hat T}_{i,e}^{\rm dur}, 
{\rm max}\!\! \left( \hat{T}_m^{\rm off}\!\! -\!\!\sum \limits_{j=1}^{i-1} {\hat T}_{j,e}^{\rm dur}, 0 \right)\!\!
\right)~~~~
\\
+ \!\!\sum \limits_{i=1}^{n_{e}^{\rm exe}(\tau_t^{\prime})}\!\!\!\! \frac{\eta}{f_e}(n_{e}^{\rm exe}(\tau_t^{\prime})\!-\!i\!+\!1)^2(L_{i,e}^{\rm sort}(\tau_t^{\prime}) \!
- \!L_{i-1,e}^{\rm sort}(\tau_t^{\prime})) + {\hat T_m}^{\rm off},
\label{eq:Reward of delay B}
\end{split}
\end{equation}
where ${\hat T_m}^{\rm off}$ is the estimated offloading delay for task $m$ with Eq.\eqref{eq:offloading delay}.
In Eq.\eqref{eq:Reward of delay B}, the first term to the right of the equation estimates the sum of delay for tasks $\mathcal{M}_{e}(\tau_t)$ from time instant $\tau_t$ to $\tau_t^{\prime}$.
The second term to the right of Eq.\eqref{eq:Reward of delay B} estimates the sum of delay for tasks $\mathcal{M}_{e}(\tau_t^{\prime})$ from time instant $\tau_t^{\prime}$ to infinity.
The expression $\frac{\eta}{f_e}(L_{i,e}^{\rm sort}(\tau_t^{\prime}) \!- \!L_{i-1,e}^{\rm sort}(\tau_t^{\prime})$ in Eq. \eqref{eq:Reward of delay B} represents the required time from completing residual size $L_{i-1,e}^{\rm sort}(\tau_t^{\prime})$ to completing residual size $L_{i,e}^{\rm sort}(\tau_t^{\prime})$. To simplify the calculation of $L_{1,e}^{\rm sort}(\tau_t^{\prime})-L_{0,e}^{\rm sort}(\tau_t^{\prime})$, we define $L_{0,e}^{\rm sort}(\tau_t^{\prime})=0$ specifically.

To summarize, if the agent offloads task $m$ to server $e$ at step $t$, the reward of delay is
\begin{equation}
\begin{split}
r_{\rm T}(\boldsymbol{s}_t,a_t) = -{\hat T_m}^{\rm off} + \sum \limits_{i=1}^{n_{e}^{\rm exe}(\tau_t)} (n_{e}^{\rm exe}(\tau_t)-i+1) {\hat T}_{i,e}^{\rm dur}
~~~~~~~~~~~~~~~~~~~~
\\
- \!\!\sum \limits_{i=1}^{n_{e}^{\rm exe}(\tau_t)} (n_{e}^{\rm exe}\!-\!i\!+1) {\rm min}\!\! 
\left( 
{\hat T}_{i,e}^{\rm dur}, 
{\rm max} \left( \hat{T}_m^{\rm off} \!\!-\!\!\sum \limits_{j=1}^{i-1} {\hat T}_{j,e}^{\rm dur}, 0\!\! \right)\!\!
\right)~~~~~~~
\\
- \sum \limits_{i=1}^{n_{e}^{\rm exe}(\tau_t^{\prime})} \frac{\eta}{f_e}(n_{e}^{\rm exe}(\tau_t^{\prime})-i+1)^2(L_{i,e}^{\rm sort}(\tau_t^{\prime}) - L_{i-1,e}^{\rm sort}(\tau_t^{\prime})).
~~~~~~~~~~~
\label{eq:Reward of delay summarize}
\end{split}
\end{equation}

To achieve the MORL algorithm, we compute a scalarized reward given preference $\boldsymbol{\omega}$:
\begin{equation}
r_{\boldsymbol{\omega} }(\boldsymbol{s}_t,a_t)
=\boldsymbol{\omega}^{T} \times (\alpha_{\rm T}{r}_{\rm T}(\boldsymbol{s}_t,a_t),\alpha_{\rm E}{r}_{\rm E}(\boldsymbol{s}_t,a_t)),
\label{fun:scalarized reward}
\end{equation}
where $\alpha_{\rm T}$ and $\alpha_{\rm E}$ are coefficients for adjusting  delay ${r}_{\rm T}(t)$ and energy consumption ${r}_{\rm E}(t)$ to the same order of magnitude. The total reward of one episode is
\begin{equation}
R_{\boldsymbol{\omega}}=\sum \limits_{t=1}^{T} r_{\boldsymbol{\omega} }(\boldsymbol{s}_t,a_t).
\label{eq:total reward}
\end{equation}

\subsection{MORL Scheduling}\label{subsec:MORL Scheduling}
We train DRL-based scheduling policies based on a PPO algorithm \cite{schulman2017proximal}, which is a family of policy gradient (PG) methods. The PPO algorithm can sample the data from the transition several times instead of one time within each episode. It improves the sampling efficiency than traditional PG methods. The neural networks with parameters $\boldsymbol{\theta}$ contain an actor network and a critic network.
In the training phase, the MORL algorithm trains a parametric network for each preference. In the evaluation phase, the parametric network evaluates the Pareto front of energy consumption and delay for multi-edge servers in the MEC environment.

We use generalized advantage estimator (GAE) technology to reduce the variance of policy gradient estimates\cite{schulman2015high}. The GAE advantage function for objective $i \in \{\rm T,\rm E \}$ is
\begin{equation}
\begin{array}{l}
\!\!\!{{\hat A}_i}(t) \!= \!
\sum\limits_{{t^\prime } = t}^{T - 1}\!\! {\gamma \lambda \left( {{\alpha _i}{r_i}({\boldsymbol{s}_{t'}},{a_{t'}})\!\! + \!\!\gamma {V_{i,{\bf{\theta }}}}({\boldsymbol{s}_{t' + 1}})\!\! -\!\! {V_{i,{\bf{\theta }}}}({\boldsymbol{s}_{t'}})} \right)} ,
\end{array}
\end{equation}
where $\lambda$ is a GAE discount factor within $[0,1]$, and $V_{i,\boldsymbol{\theta}}(\boldsymbol{s}(t))$ denotes the value of state $\boldsymbol{s}(t)$. 
Value function $V_{i,\boldsymbol{\theta}}(\cdot)$ is estimated by a critic network.

In the PPO algorithm, the gradient direction of objective $i \in \{\rm T,\rm E\}$ is given as

\begin{equation}
\begin{array}{l}
\nabla_{\boldsymbol{\theta}} L_i^{\rm {clip}}(\boldsymbol{\theta})\!=\!\mathbb{E}_t [\min\left( r^{\rm pr}_t(\boldsymbol{\theta}), clip(r^{\rm pr}_t(\boldsymbol{\theta}), 1\!-\!\epsilon, 1\!+\!\epsilon) \right)
\\~~~~~~~~~~~~~~~~~ \hat{A}_i(t) \nabla \log \pi_{\boldsymbol{\theta}}(a_t|\boldsymbol{s}_t)  ],
\end{array}
\end{equation}
where $\epsilon$ is a clip hyperparameter. The probability ratio is $r_t^{\rm pr}(\boldsymbol{\theta})=\frac{\pi_{\boldsymbol{\theta}}(a_t|\boldsymbol{s}_t)}{\pi_{\boldsymbol{\theta}_{old}}(a_t|\boldsymbol{s}_t)}$. The surrogate objective is $r_t^{\rm pr}(\boldsymbol{\theta})\hat{A}_t$, which corresponds to a conservative policy iteration. The objective is constrained by $clip(r_t^{\rm pr}(\boldsymbol{\theta})\hat{A}_t, 1-\epsilon, 1+\epsilon)$, to penalize the policy move outside interval $[1-\epsilon, 1+\epsilon]$.

\begin{figure}[t]
\centering
\includegraphics[width=55mm]{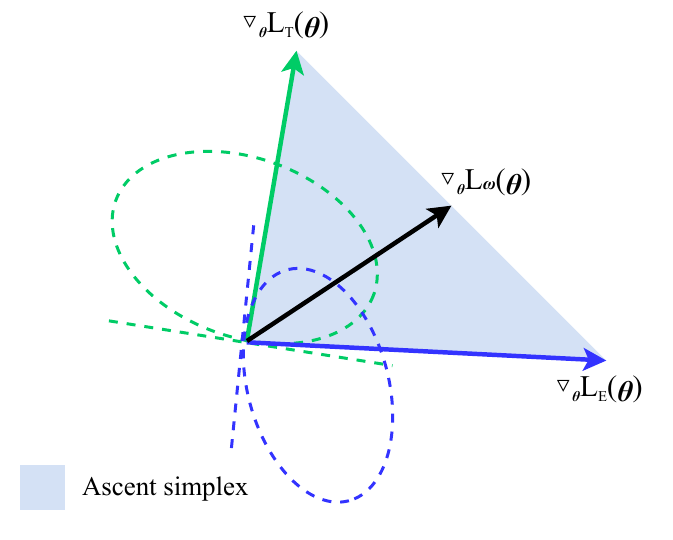}
\caption{The ascent simplex in a 2–objectives problem.}
\label{fig:ascent simplex}
\end{figure}

Given the gradient directions of the two objectives, a policy can reach the Pareto front by following a direction in ascent simplex \cite{parisi2014policy}. An ascent simplex is deﬁned by the convex combination of single–objective gradients. As shown in Fig. \ref{fig:ascent simplex}, the green arrow and blue arrow denote the gradient directions of the delay objective and energy consumption objective, respectively. The light blue area stands for an ascent simplex.

For  reward function $r_{\boldsymbol{\omega}}(\cdot)$, the gradient direction of preference $\boldsymbol{\omega}$ is 
\begin{equation}
\begin{array}{l}
\nabla_{\boldsymbol{\theta}} L_{\boldsymbol{\omega}}^{\rm {clip}}(\boldsymbol{\theta})\!=\!\mathbb{E}_t [\min\left( r_t^{\rm pr}(\boldsymbol{\theta}), clip(r_t^{\rm pr}(\boldsymbol{\theta}), 1\!-\!\epsilon, 1\!+\!\epsilon) \right)
\\~~~~~~~~~~~~~~~~~~~ \boldsymbol{\omega}^{\mathrm{T}}(\hat{A}_{1}(t),\hat{A}_{2}(t)) \nabla \log \pi_{\boldsymbol{\theta}}(a_t|{\boldsymbol{s}_t})  ]\\
~~~~~~~~~~~~~~~=\boldsymbol{\omega}^{\mathrm{T}}(\nabla_{\boldsymbol{\theta}} L_{1}^{\rm {clip}}(\boldsymbol{\theta}), \nabla_{\boldsymbol{\theta}} L_{2}^{\rm {clip}}(\boldsymbol{\theta})).
\end{array}
\end{equation}
The vector $\nabla_{\boldsymbol{\theta}} L_{\boldsymbol{\omega}}^{\rm {clip}}(\boldsymbol{\theta})$ is a gradient direction in ascent simplex. It makes a policy to the Pareto front by optimizing neural network parameters $\boldsymbol{\theta}$. 

\begin{figure}[t]
        \small
        \centering
        \includegraphics*[width=80mm]{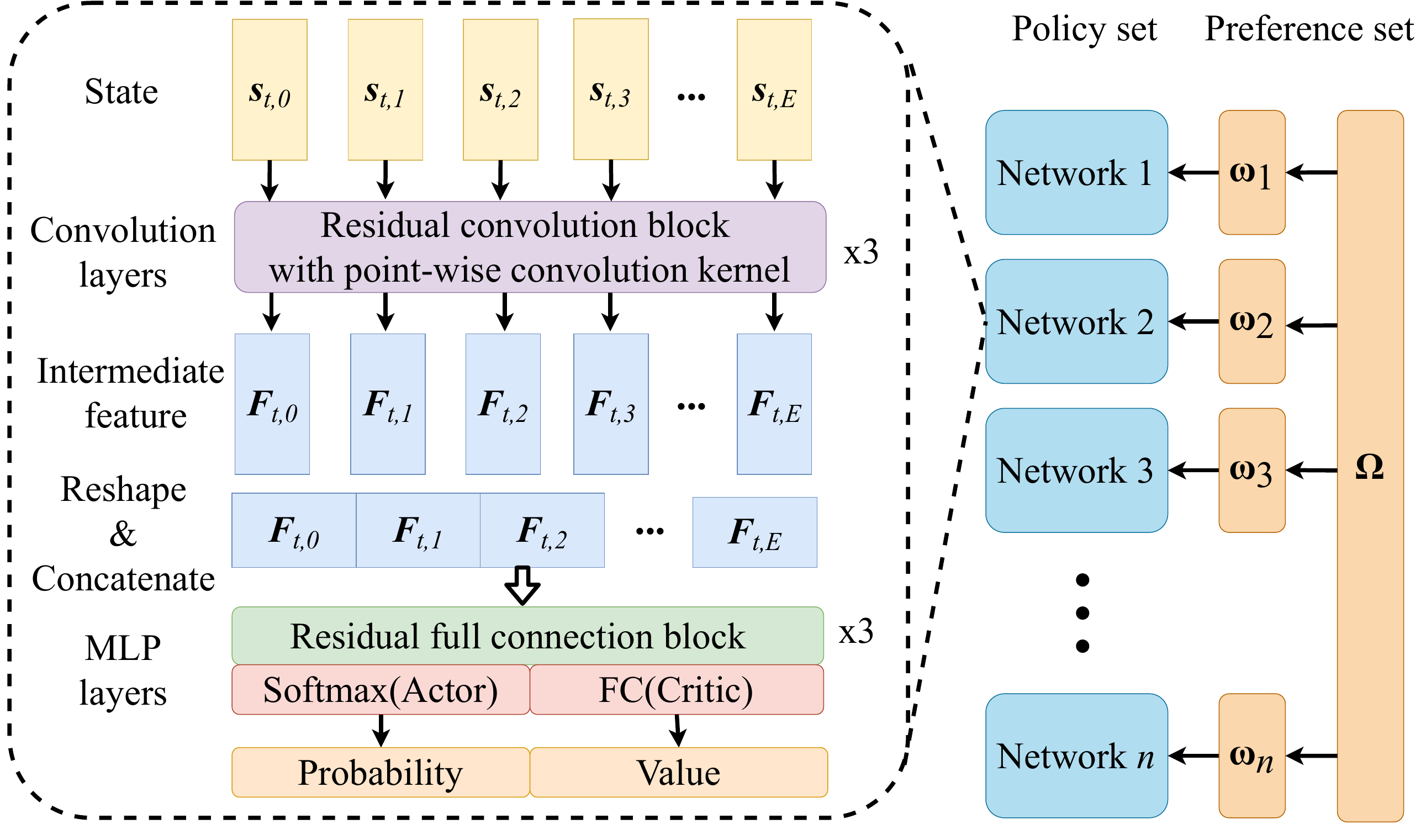}
        \caption{The neural network framework of a scheduling policy.}
        \label{fig:Network}
\end{figure}

As an example shown in Fig. \ref{fig:Network}, a neural network contains convolution layers and multi-layer perceptron (MLP) layers. The convolution layers encode the input state with point-wise convolution kernel and turn information vector $\boldsymbol{s}_t$ of each server to feature vector $\boldsymbol{F}$. We reshape all feature vectors and concatenate them to get the total feature vector. The MLP layers encode the total feature vector to get the output. For an actor-network, the output is probability $\pi_{\boldsymbol{\theta}}(a_t|\boldsymbol{s}_t)$ of each action. For a critic network, the output is estimated value $\boldsymbol{\omega}^{\rm T}[V_{{\rm T},\boldsymbol{\theta}}(\boldsymbol{s}_t),V_{\rm E,\boldsymbol{\theta}}(\boldsymbol{s}_t)]$ for preference $\boldsymbol{\omega}$. 
Additionally, we apply deep residual learning technology \cite{he2016deep} to build the neural network architecture to address the problem of vanishing/exploding gradients.

We present the proposed MORL algorithm in Algorithm 1. For each preference ${\boldsymbol{\omega} }$ in set $\Omega$, we train a policy with PPO method to maximize reward $R_{\boldsymbol{\omega} }$ and approximate Pareto front $PF(\Pi)$. To improve the training efficiency achieved by \cite{natarajan2005dynamic}, we reuse trained neural network parameters $\boldsymbol{\theta}_{\boldsymbol{\omega}_i}$ ($i\in \{1,2,…,n-1\}$) to initialize the next parameters $\boldsymbol{\theta}_{\boldsymbol{\omega}_{i+1}}$, with a similar preference.

\begin{algorithm}
\caption{MORL-based Scheduling}
\begin{algorithmic}[1]
\STATE Initialize replay memory buffer $D_{\boldsymbol \omega}$, policy parameters $\boldsymbol{\theta}_{\boldsymbol{\omega}}$ for each preference $\boldsymbol \omega$ 
\STATE Initialize the learning rate $\alpha$ and the number of episodes $T^{\rm epi}$ for training.
\STATE Set policies set $\Pi \gets\emptyset$
\FOR{each preference $\boldsymbol \omega$}
	\FOR{each episode $T^{\rm epi}$}
        \FOR{each step $t$}
                \STATE $a_t \sim {\pi_{\boldsymbol{\theta}_{\boldsymbol{\omega}}}}(\boldsymbol{s}_t)$ 
                \STATE $\boldsymbol{s}_{t+1}\sim \mathcal{T}(\boldsymbol{s}_{t+1}|\boldsymbol{s}_t,)$
                \STATE$D_{\boldsymbol \omega} \!=\! D_{\boldsymbol \omega}\cup\{(\boldsymbol{s}_t, a_t, r_{\boldsymbol \omega}(\boldsymbol{s}_t, a_t), \boldsymbol{s}_{t+1}\}$
         \ENDFOR
         \STATE $\boldsymbol{\theta}_{\boldsymbol{\omega}} \gets \boldsymbol{\theta}_{\boldsymbol{\omega}} + \alpha\nabla_{\boldsymbol{\theta_{\omega}}} L_{\boldsymbol{\omega}}^{clip}(\boldsymbol{\theta_{\omega}})$
    \ENDFOR
    \STATE $\Pi \gets \Pi \cup {\pi_{\boldsymbol {\boldsymbol{\theta}_{\omega}} }}$
\ENDFOR
\STATE Compute Pareto front $PF(\Pi)$
\end{algorithmic}
\end{algorithm}

\section{Simulation Results}
In this section, we evaluate the performances of the MORL scheduling scheme and compare it with benchmarks. We introduce the simulation setup and evaluation metrics. Then, we analyze the Pareto fronts and compare them with the benchmarks.

\subsection{Simulation Setup}
We set the preference set as $\Omega$ with an equal interval $0.02$ and obtain $50$ preferences to fit the Pareto front. Each preference's performance contains total delay and energy consumption for all tasks in one episode. We evaluate a performance (delay or energy consumption) with an average of $1000$ episodes. 
Furthermore, we analyze the Pareto front of the proposed scheme and compare it with benchmarks.
A disk coverage has a radius of $1000$m to $2000$m for a cloud server and $50$m to $500$m for an edge server. Each episode needs to initial different radiuses for the cloud and edge servers.
We set the mean of task size $\bar{L}$ according to Eq. \eqref{eq:Balance}.

\begin{table}[t]
\vspace{0.05in}
\caption{Model Parameters}
\begin{center}
\begin{spacing}{1.07}
\begin{tabular}{m{50mm}|m{22mm}}
\hline\hline
{\bf Resource Scheduling Hyperparameters} & {\bf Values}\\
\hline
The number of steps for one episode $T$ &  $100$\\
\hline
Step duration  $\Delta t$  & $1~\rm{s}$ \\
\hline
The number of users $U$ & $10$\\
\hline
The number of tasks $M$ & $100$\\
\hline
System bandwidth $W$ & $16.6$MHz\cite{815305}\\
\hline
Offloading power ${p^{\rm off}}$ & $10~\rm{mW}$ \\
\hline
The number of CPU cycles $\eta$ for one-bit task & $10^{3}$\\
\hline
Effective capacitance coefﬁcient $\kappa$ & $5 \times 10^{-31}$ \\
\hline
CPU frequency of cloud server $f_0$& $4.0 ~\rm{GHz}$ \\
\hline
CPU frequency of edge server $f_e$& $2.0 ~\rm{GHz}$ \\
\hline
Poisson arrival rate $\lambda_ p$ for each user & $0.1$ \\
\hline
{\bf DRL Hyperparameters} & {\bf Values}\\
\hline
The episodes for training $T^{\rm epi}$ &  $1.92 \times 10^6$\\
\hline
Replay memory & $1 \times 10^5$ \\
\hline
Batch size & $4096$ \\
\hline
Learning rate & $1\times10^{-6}$ \\
\hline
Discount factor $\gamma$ & $0.9$ \\
\hline
GAE  discount factor $\lambda$ & $0.95$\\
\hline
Clip parameter $\epsilon$ & $0.2$\\
\hline\hline
\end{tabular}
\end{spacing}
\end{center}
\label{table:1}
\end{table}

\begin{figure}[t]
\centering
\subfloat[]{
\includegraphics[width=40mm]{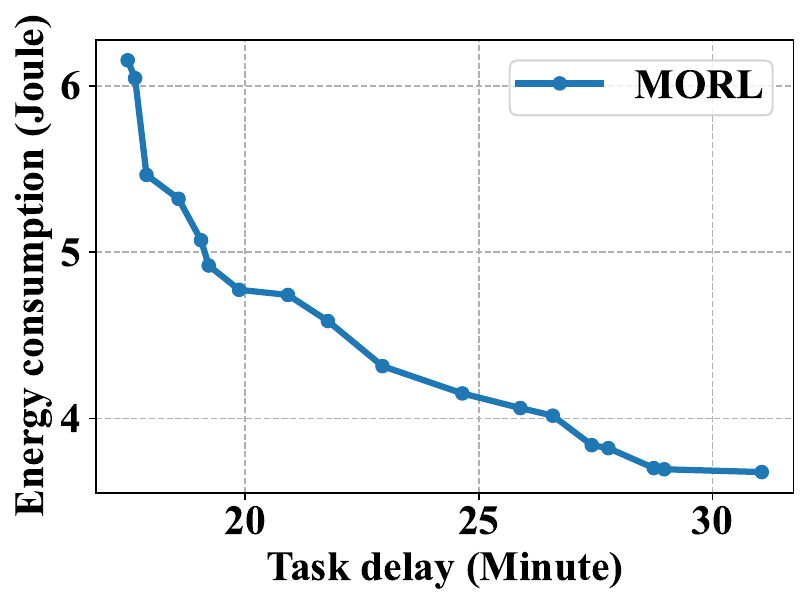}
\label{fig:Pareto front1}
}
\hspace{-2mm}
\subfloat[]{
\includegraphics[width=40mm]{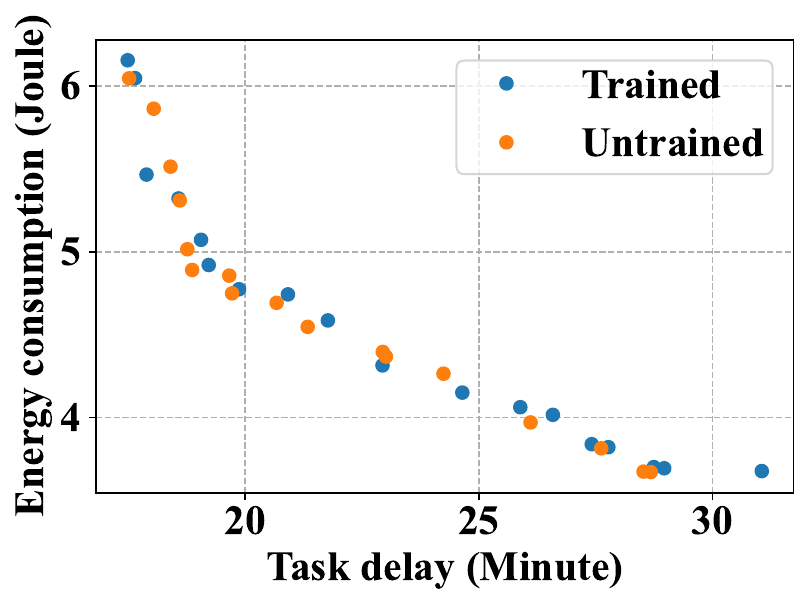}
\label{fig:Pareto front2}
}
\caption{The Pareto front of the MORL scheme.}
\end{figure}


\begin{figure}[t]
\centering
\includegraphics[width=60mm]{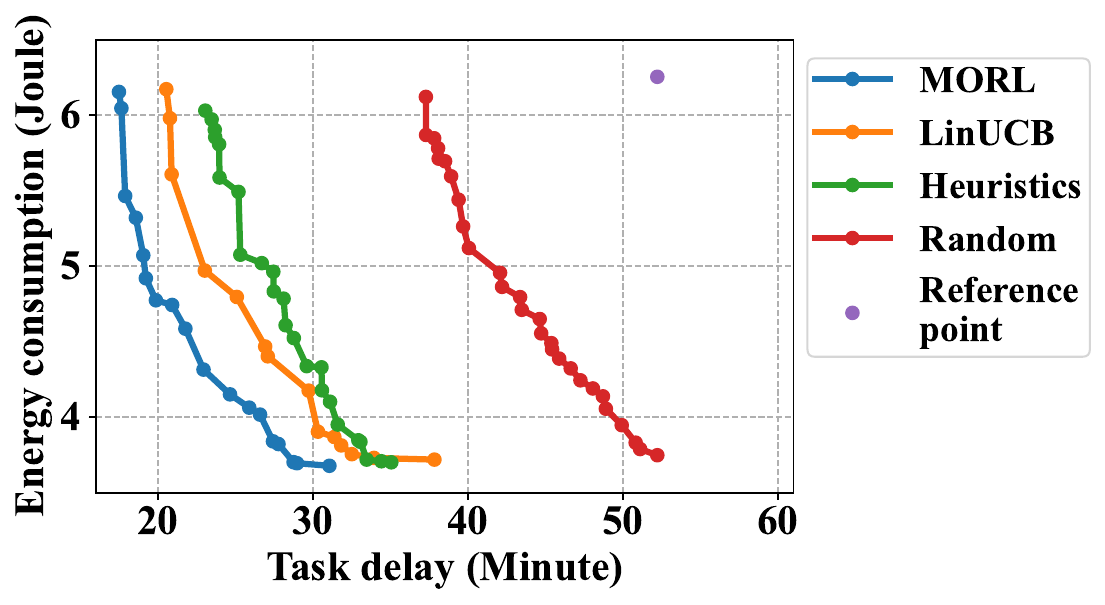}
\caption{The Pareto fronts of MORL scheme and other schemes.}
\label{fig:Base line}
\end{figure}
\vspace{0pt}
\subsection{Evaluation Metrics}
We consider the following metrics to evaluate the performances of the proposed algorithms.

\begin{itemize}
\item \textbf{Energy Consumption:} The total energy consumption of one episode given as 
$\textstyle \sum \limits_{m=1}^{M} E_{m}^{\rm off} + E_{m}^{\rm exe}$, and average energy consumption per Mbits task of one episode given by $\textstyle \sum \limits_{m=1}^{M} \frac{E_{m}^{\rm off} + E_{m}^{\rm exe}}{M\bar{L}}$.

\item \textbf{Task Delay:} The total task delay given as $\textstyle \sum \limits_{m=1}^{M} T_{m}^{\rm off} + T_{m}^{\rm exe}$ and average delay per Mbits task of one episode given by $\textstyle \sum \limits_{m=1}^{M} \frac{T_{m}^{\rm off} + T_{m}^{\rm exe}}{M\bar{L}}$.

\item \textbf{Pareto Front:} ${PF(\Pi)\!\!=\!\!\{\pi \!\in\! \Pi~\!|\!~\nexists \pi^{\prime}\in\Pi:\boldsymbol{y}^{\pi^{\prime}} \!\!\!\succ_P\!\! \boldsymbol{y}^{\pi}} \}$, where the symbols are deﬁned by Eq. \eqref{def:Pareto front}.

\item \textbf{Hypervolume metric:}

$\mathcal{V}(PF(\Pi))=\int_{\mathbb{R}^2} {\mathbb{I}_{V_h(PF(\Pi))}(z)dz}$, where the symbols are deﬁned by Eq. \eqref{def:Hypervolume metric}.
 
\end{itemize} 

\subsection{Simulation results}
\subsubsection{Pareto Front Analysis} Fig. \ref{fig:Pareto front1} presents the Pareto front of the proposed MORL scheme. In this scenario, the number of edge servers is $E=8$, and the mean of task size $\bar L=20~\rm{Mbits}$. The Pareto front shows that minimizing the delay (the leftmost point) increases energy by $67.3\%$, but minimizing energy (the rightmost point) increases the delay by $77.6\%$. Fig. \ref{fig:Pareto front2} shows the points of the Pareto front with trained and untrained preferences. Each untrained preference lies intermediate to the adjacent trained preferences. The result shows that by reusing trained parameters to the most similar preference, our MORL scheme has generalization for new preferences.

\subsubsection{Performance Comparison with Benchmarks}
We evaluate the performance of the proposed MORL algorithms and compare it with a linear upper confidence bound (LinUCB)-based scheme \cite{li2010contextual}, a heuristics-based scheme, and a random-based scheme. LinUCB algorithms belong to contextual multi-arm bandit (MAB) algorithms, widely used in task offloading problems \cite{chen2019task,zhao2022collaboration}. Some work \cite{bi2018computation,tran2018joint
} apply heuristic methods to schedule for offloading.
\begin{itemize}
\item \textit {LinUCB-based scheme}: Offloading scheme based on a kind of contextual MAB algorithm. This scheme uses states as MAB contexts and learns a policy by exploring different actions.

\item \textit {Heuristics-based scheme}: Heuristic methods greedily select the server with the optimal weighted sum of estimated running speed and energy consumption for the current step.

\item \textit {Random-based scheme}: The agent offloads a task to a cloud server or a random edge server according to probability. We adjust the probability to compute a Pareto front.
 
\end{itemize}

Fig. \ref{fig:Base line} illustrates the Pareto front comparison of the proposed MORL scheme with other schemes. In this scenario, the system has $E = 8$ and $\bar L=20~\rm{Mbits}$. 
We select the position which denotes the maximum delay and energy consumption of the performance profiles in Fig. \ref{fig:Base line} as the reference point to compute the hypervolumes. The hypervolume of the proposed MORL scheme is $80.7$, the LinUCB-based scheme is $69.9$, the heuristics-based scheme is $63.9$, and a random-based scheme is $24.2$. Compared with a LinUCB-based scheme and a random-based scheme, the proposed MORL scheme increases the hypervolume of the Pareto front by $\textstyle \frac{80.7-69.9}{69.9}=15.5\%$ and $\textstyle \frac{80.7-24.2}{24.2}=233.1\%$. As shown, the proposed MORL scheme signiﬁcantly outperforms other schemes. The MORL scheme has dynamic adaptability to learn the dynamics of task arrival and server load, which enables it to achieve better scheduling. 
\begin{figure}[t]
\centering
\subfloat[Pareto fronts of total delay and energy consumption]{
\includegraphics[width=38mm]{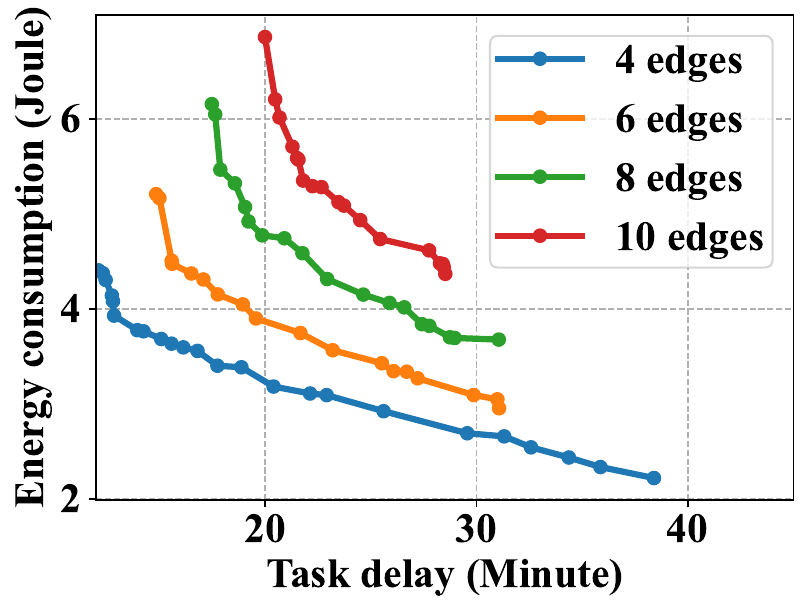}
\label{fig:Multi_edge1}
}
\hspace{0mm}
\subfloat[Pareto fronts of total delay and energy consumption per Mbits task]
{\includegraphics[width=39.5mm]{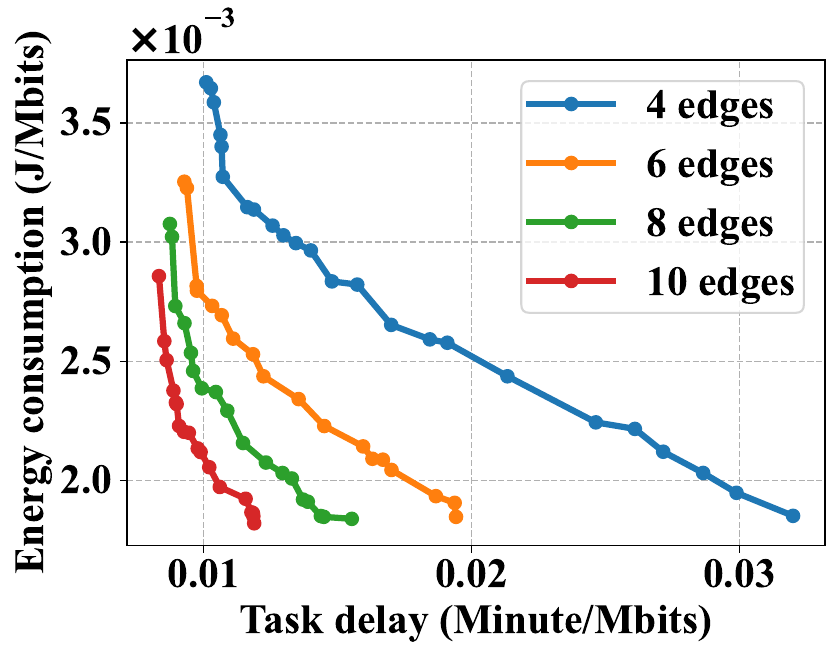}
\label{fig:Multi_edge2}
~~~~
}
\caption{Pareto fronts of the proposed MORL algorithm. 
}
\end{figure}

\subsubsection{Pareto Front Analysis in Multi-edge Scenarios}
We evaluate the Pareto front of the proposed MORL algorithm in scenarios with different edge server quantities.
Fig. \ref{fig:Multi_edge1} illustrates the Pareto fronts of the proposed MORL algorithm in the case of edge quantity $E \in \{4,6,8,10\}$. The mean of task size, represented by $\bar{L}$, is determined by Eq. \eqref{eq:Balance} to balance the supply and demand of computational capability. 
The result shows that, in the balance case, the Pareto front of fewer edge servers and less demand case can dominate the more one. It means that while more edge servers may increase computational capability, matching them with more task demands may result in increased total energy consumption and task delay. The performances are computed per $1$ Mbits task in Fig. \ref{fig:Multi_edge2} for a fair comparison. As the number of edge servers increases, the Pareto front of a more edge servers case can dominate the less one. 
The result shows that though more edge servers match more task demands, deploying more edge servers can significantly improve delay and energy consumption per Mbits tasks for each preference.

\section{Conclusion}
In this work, we investigated the offloading problem in MEC systems and proposed a MORL-based algorithm that can achieve Pareto fronts. A key advantage of the proposed MORL method is that it employs a MORL framework to offload tasks adopting various preferences, even untrained preferences.

We present a novel MOMDP framework for the multi-objective offloading problem in MEC systems. Our framework includes two key components: (1) a well-designed encoding method to construct features of multi-edge MEC systems. (2) a sophisticated reward function to evaluate the immediate utility of delay and energy consumption. Simulation results demonstrate the effectiveness of our proposed MORL scheme, which achieves Pareto fronts in various scenarios and outperforms benchmarks by up to $233.1\%$.

\section*{Acknowledgments}
The research leading to these results received funding from “Research on Combinatorial Optimization Problem Based on Reinforcement Learning” supported by Beijing Municipal Natural Science Foundation under Grant Agreement Grant No. 4224092. This work was supported in part by the National Natural Science Foundation of China under Grants 62202427 and Grants 62202214. In addition, it received funding from National Key R$\&$D Program of China (2022ZD0116402). 

\bibliographystyle{IEEEtran}
\bibliography{references}
\end{document}